\documentclass[prb,showpacs,floatfix,twocolumn]{revtex4}
\usepackage{amssymb}
\usepackage{txfonts}
\usepackage{graphicx}
\usepackage{dcolumn}
\usepackage{float}
\begin{document}



\title{Two successive field-induced spin-flop transitions in
 single-crystalline CaCo$_{2}$As$_{2}$ }
\author{B. Cheng}
\author{B. F. Hu}
\author{R. H. Yuan}
\author{T. Dong}
\author{A. F. Fang}
\author{Z. G. Chen}
\author{G. Xu}
\author{Y. G. Shi}
\author{P. Zheng}
\author{J. L. Luo}
\author{N. L. Wang}

\affiliation{ Institute of Physics, Chinese Academy of Sciences,
Beijing 100080, People's Republic of China}
%


\begin{abstract}
CaCo$_{2}$As$_{2}$, a ThCr$_{2}$Si$_{2}$-structure compound,
undergoes an antiferromagnetic transition at \emph{T$_{N}$}=76K with
the magnetic moments being aligned parallel to the \emph{c} axis.
Electronic transport measurement reveals that the coupling between
conducting carriers and magnetic order in CaCo$_{2}$As$_{2}$ is much
weaker comparing to the parent compounds of iron pnictide. Applying
magnetic field along \emph{c} axis induces two successive spin-flop
transitions in its magnetic state. The magnetization saturation
behaviors with \emph{\textbf{H}$\parallel$c} and
\emph{\textbf{H}$\parallel$ab} at 10K indicate that the
antiferromagnetic coupling along \emph{c} direction is very weak.
The interlayer antiferromagntic coupling constant \emph{J$_{c}$} is
estimated to be about 2 meV.

\end{abstract}


\maketitle

\section{\label{sec:level2}INTRODUCTION}

Magnetic responses of magnetic materials to external fields have
been one of the most active fields in condensed matter physics due
to their enormous value for fundamental researches and practical
applications. Competition between exchange energy,
magnetocrystalline anisotropy energy and Zeeman energy could
introduce many fascinating magnetic phenomena in magnetic
materials.\cite{RMP1,LiLun1} Especially, in an antiferromagnet with
low anisotropy, a magnetic field applied parallel to the easy axis
could induce a transition to a phase in which the magnetic moments
lie in a direction perpendicular to the external magnetic field.
This is the so-called spin-flop transition. Spin-flop phenomena have
been observed in many magnetic materials,\cite{3D1,3D2-LSMO,3D3,3D4}
including many low dimensional
antiferromagnets.\cite{1D1,1D2,1D3,2D1}

Recently, the discovery of hign-\emph{T$_{c}$} iron pnictide
superconductors opens a playground for the community to explore the
magnetism and its interplay with superconductivity.\cite{Fe1} Most
parent compounds of FeAs-based superconductors exhibit a long-range
antiferromagnetic (AFM) order at low temperature. Similar to the
cuprates, hole or electron doping will suppress AFM and introduce
superconductivity.\cite{Fe2,Fe3} In 1111 family, \emph{T$_c$}
reaches up to 55K,\cite{Fe4} which is much higher than the value
expected from the traditional electron-phonon coupling theory. It is
widely believed that the superconductivity in iron pnictides has an
unconventional origin.

 Moderate hole or electron doping will destroy the
long-range antiferromagnetic order in iron pnictide parents.
However, a complete substitution of Fe by Co in ReFeAsO (Re=La-Gd)
will introduce complex magnetic phenomena. LaCoAsO shows
ferromagnetic (FM) order below 55K with saturation moment of 0.3
$\sim$ 0.4$\mu$$_B$ per Co atom.\cite{Co1} In ReCoAsO (Re=Ce-Gd),
the existence of \emph{4f} electrons in rare-earth elements leads to
extra complexity in magnetism.\cite{Co1,Co2} Recent neutron
diffraction experiments reveal NdCoAsO undergoes three magnetic
transitions: (a) ferromagnetic transition at 69K from Co moments,
(b) transition from FM to AFM at 14K and (c) antiferromagnetic order
of Nd 4\emph{f} moments below 1.4K.\cite{Co3,Co4} Neutron
experiments indicate that all ordered moments lie in the \emph{ab}
plane. The moments on Co atoms in each CoAs layer are
ferromagnetically ordered, and these layers are aligned
antiferromagnetically along \emph{c} direction. The two Nd sites in
each NdO layer are aligned antiferromagnetically and alternate in
direction between layers. The interplay between 3\emph{d} and
4\emph{f} electrons may play an important role in these successive
magnetic transitions.

However, BaCo$_{2}$As$_{2}$, which belongs to 122 family, exhibits
paramagnetic behavior above 1.8K.\cite{Co5} The enhancement of
susceptibilities relative to the weak correlated electron systems
indicates that BaCo$_{2}$As$_{2}$ is close to a magnetic quantum
critical point. In this article, we report our exploration of
single-crystalline CaCo$_{2}$As$_{2}$. Different from
BaCo$_{2}$As$_{2}$, we found that CaCo$_{2}$As$_{2}$ undergoes an
antiferromagnetic transition at 76K with magnetic moments being
aligned parallel to the \emph{c} axis. Interestingly, applying
magnetic field parallel to the easy axis induces two successive
spin-flop transitions in its magnetic state. Our studies indicate
that the magnetic coupling between \emph{ab} plane is very weak, so
that the magnetic ground state of CaCo$_{2}$As$_{2}$ can be
disturbed easily by a moderate external magnetic field.

\section{\label{sec:level2}EXPERIMENTAL DETAIL}

\begin{figure}[b]
\scalebox{0.45} {\includegraphics [bb=330 30 8cm 12cm]{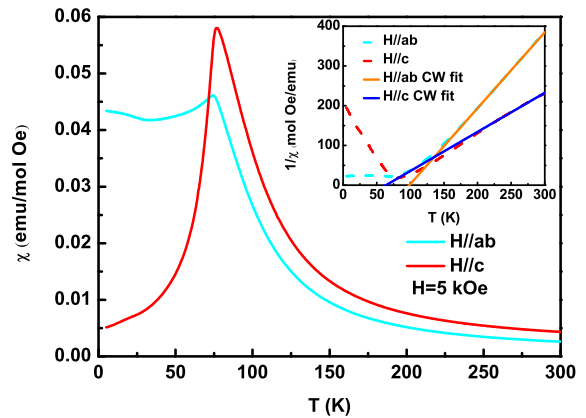}}
\caption{(color online) Temperature dependent magnetic
susceptibilities along \emph{ab} plane and \emph{c} axis in 5 kOe.
The inset is the Curie-Weiss fits to the high temperature parts of
susceptibilities.}
\end{figure}

CaCo$_{2}$As$_{2}$ single crystals were grown by a self-flux method
similar to the procedures described in many
references.\cite{Co5,flux} Typical crystal size was $\sim$ 5
$\times$ 5 $\times$ 0.1 mm$^{3}$. Resistivity and specific heat
measurements were performed on Quantum Design physical property
measurement system (PPMS). Dc magnetization was measured as
functions of temperature and magnetic field using Quantum Design
instrument superconducting quantum interference device (SQUIT-VSM)
and PPMS.

\section{\label{sec:level2}RESULTS AND DISCUSSIONS}

\begin{figure}[b]
\scalebox{0.45} {\includegraphics [bb=600 20 -1cm 13cm]{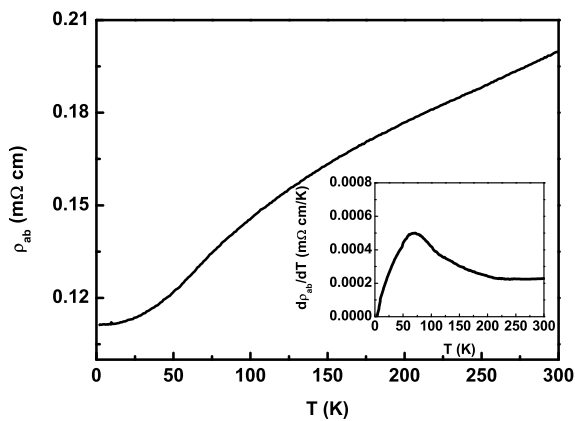}}
\caption{Temperature dependent resistivity of CaCo$_{2}$As$_{2}$ in
zero field with \emph{\textbf{I}$\parallel$ab}. The inset is the
derivative \emph{d$\rho$$_{ab}$/dT} as a function of temperature. }
\end{figure}

Figure 1 shows susceptibilities in 5 kOe along the \emph{c} axis and
\emph{ab} plane. The susceptibility for
\emph{\textbf{H}$\parallel$c} exhibits a sharp peak at
\emph{T$_{N}$} = 76K and drops rapidly with decreasing temperature,
indicative of an antiferromagnetic transition with the magnetic
moments being aligned parallel to \emph{c} axis.\cite{3D2-LSMO}
Susceptibility for \emph{\textbf{H}$\parallel$ab} shows a peak
around 76K too, but it does not decrease rapidly as
\emph{\textbf{H}$\parallel$c} and shows nearly no temperature
dependence below \emph{T$_{N}$}. Above 150K, the susceptibilities
follow the Curie-Weiss law very well. Inset of Fig. 1 shows
Curie-Weiss fits to the high temperature parts of susceptibilities.
These fits give Weiss temperature \emph{$\theta$$_{ab}$}=98K for
\emph{\textbf{H}$\parallel$ab} and \emph{$\theta$$_{c}$}=65K for
\emph{\textbf{H}$\parallel$c}. The effective moment of Co is
calculated to be 1.0\emph{$\mu$$_{B}$} for
\emph{\textbf{H}$\parallel$ab} and 1.4\emph{$\mu$$_{B}$} for
\emph{\textbf{H}$\parallel$c}. Positive Weiss temperature generally
indicates ferromagnetic coupling between the moments. We notice that
many compounds with crystal structure similar to CaCo$_{2}$As$_{2}$
show FM ordering in the basal planes, such as CaCo$_{2}$P$_{2}$,
LaCo$_{2}$P$_{2}$, CeCo$_{2}$P$_{2}$, PrCo$_{2}$P$_{2}$ and
NdCo$_{2}$P$_{2}$.\cite{P1,P2} Especially, local-density
approximation calculation (LDA) reveals that BaCo$_{2}$As$_{2}$
displays a in-plane ferromagnetic correlation even if it does not
exhibit magnetic order above 1.8K.\cite{Co5} So it is reasonable to
infer that the moments of Co atoms are ordered ferromagnetically
within \emph{ab} plane in its magnetic state. However, the magnetic
structure of CaCo$_{2}$As$_{2}$ can not been determined exactly by
the static susceptibilities. The simplest supposition of the
magnetic structure of CaCo$_{2}$As$_{2}$ is an A-type
antiferromagnetism as shown in Fig. 6(a), which means the magnetic
moments of Co atoms are aligned antiferromagnetically along the c
axis with the stacking sequence + - + -. Another possible type of
stacking sequence along c axis is + + - - as shown in Fig. 6(b).
This type of stacking sequence has been observed in
PrCo$_{2}$P$_{2}$ and NdCo$_{2}$P$_{2}$.\cite{P1,P2}  Neutron
experiments are needed to determine the exact magnetic structure of
CaCo$_{2}$As$_{2}$.

\begin{figure}[t]
\scalebox{0.30} {\includegraphics [bb=600 50 8cm 19cm]{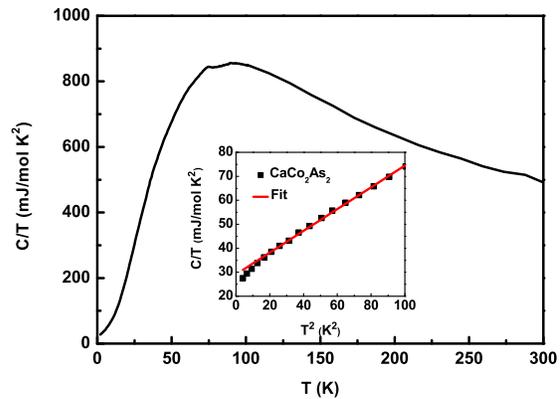}}
\caption{(color online) Temperature dependence of the specific heat
of CaCo$_{2}$As$_{2}$ plotted as \emph{C/T} vs \emph{T}. The inset
is a linear fit to the data below 10K.}
\end{figure}

Figure 2 shows the resistivity for \emph{\textbf{I}$\parallel$ab} as
a function of temperature. \emph{$\rho$$_{ab}$} decreases with
decreasing temperature, revealing a metallic behavior. There is no
clear anomaly in resistivity curve across antiferromagntic
transition temperature. Only a broad peak could be seen in the
derivative plot \emph{d$\rho$$_{ab}$/dT} as shown in the inset of
the Fig. 2. This is very different from the electronic transport
behaviors of parent compounds of iron-based superconductors, where
clear anomalies were observed in resistivity at transition
temperature.\cite{Fe2,Fe3} Absence of similar anomaly in
\emph{$\rho$$_{ab}$} of CaCo$_{2}$As$_{2}$ indicates that the
coupling between the conducting carriers and antiferromagnetical
order in CaCo$_{2}$As$_{2}$ is much weaker than that of iron
pnictide parents.

\begin{figure*}[t]

\includegraphics[clip,width=2.3in]{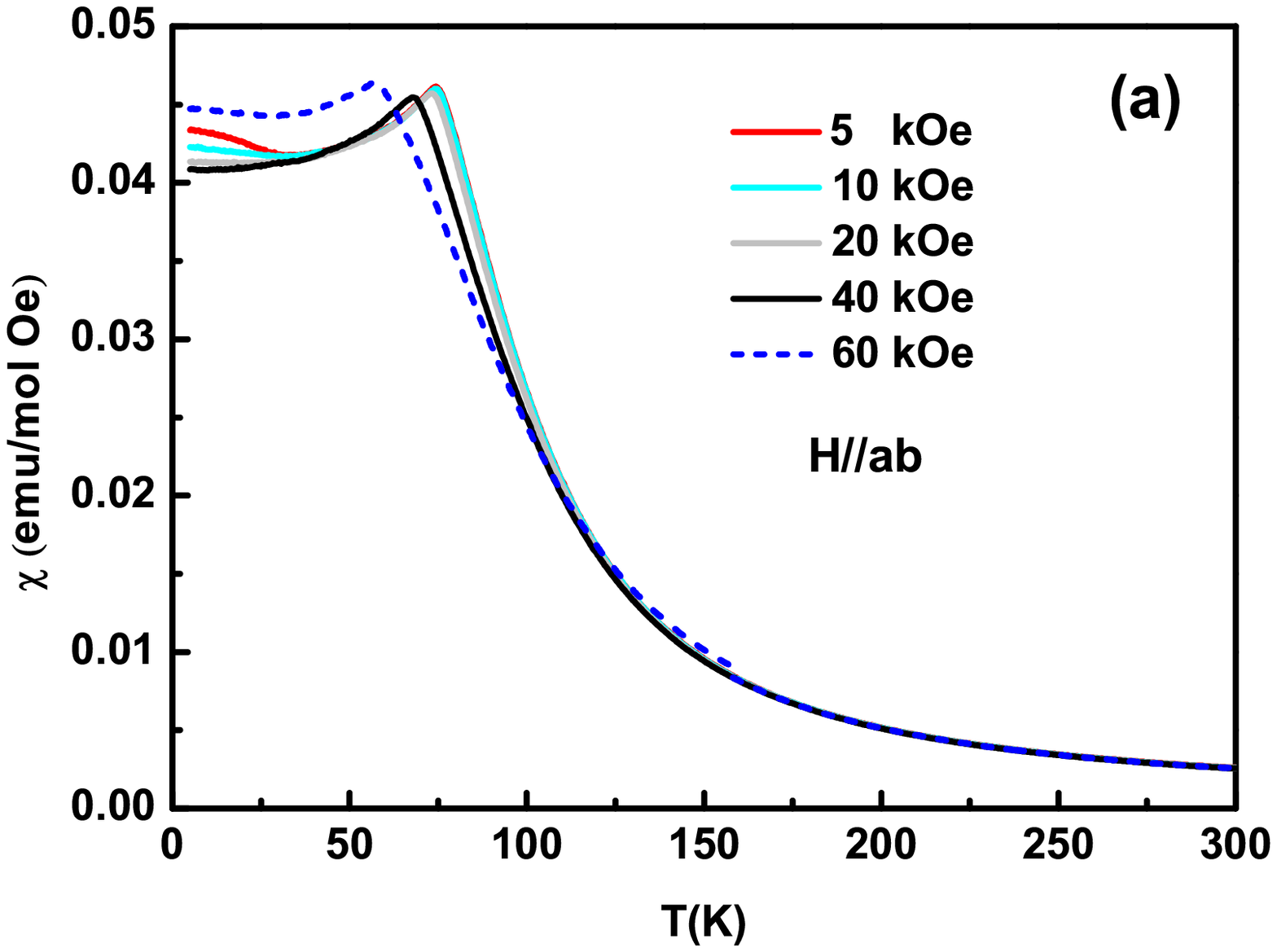}
\includegraphics[clip,width=2.3in]{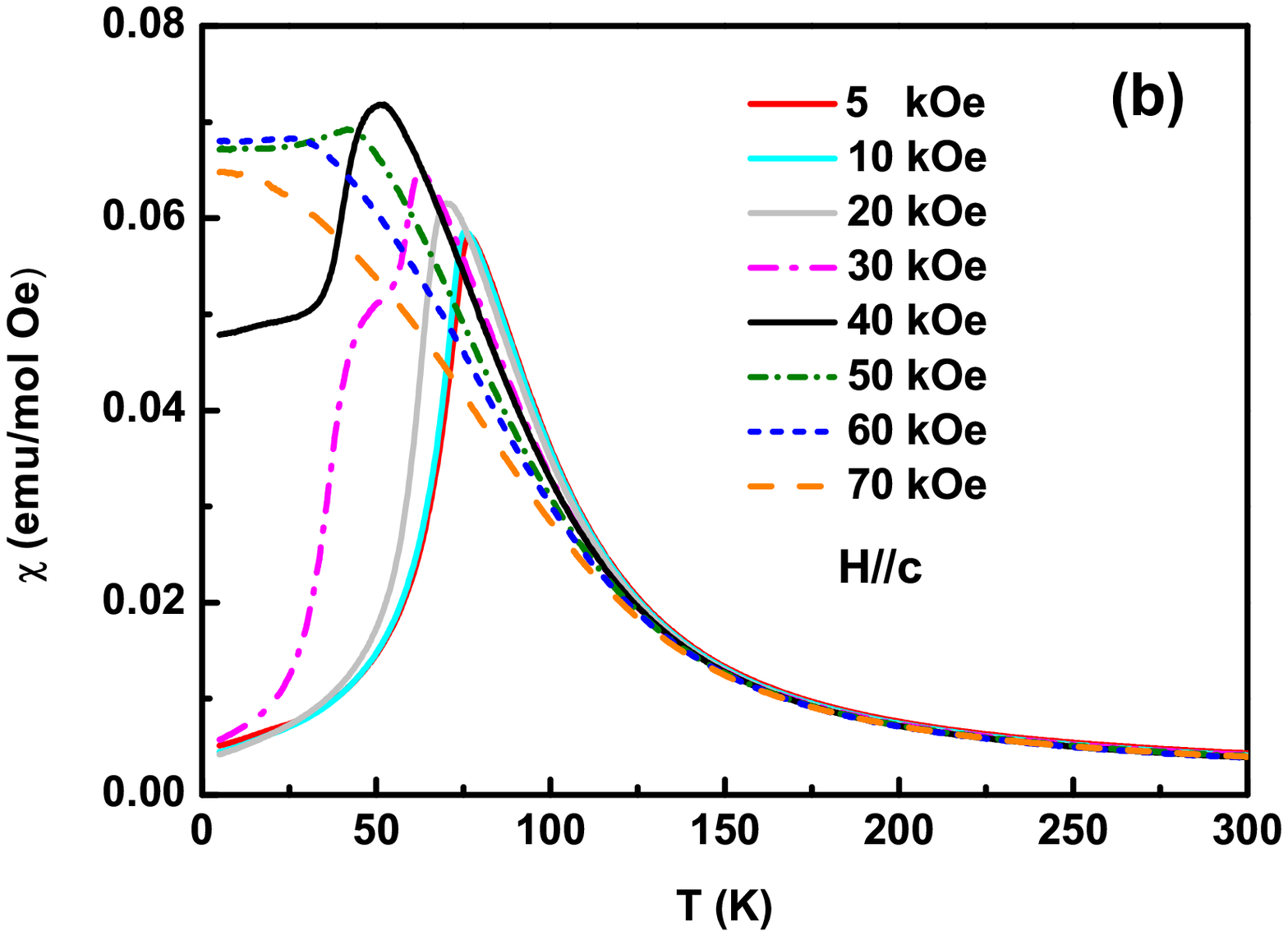}
\includegraphics[clip,width=2.3in]{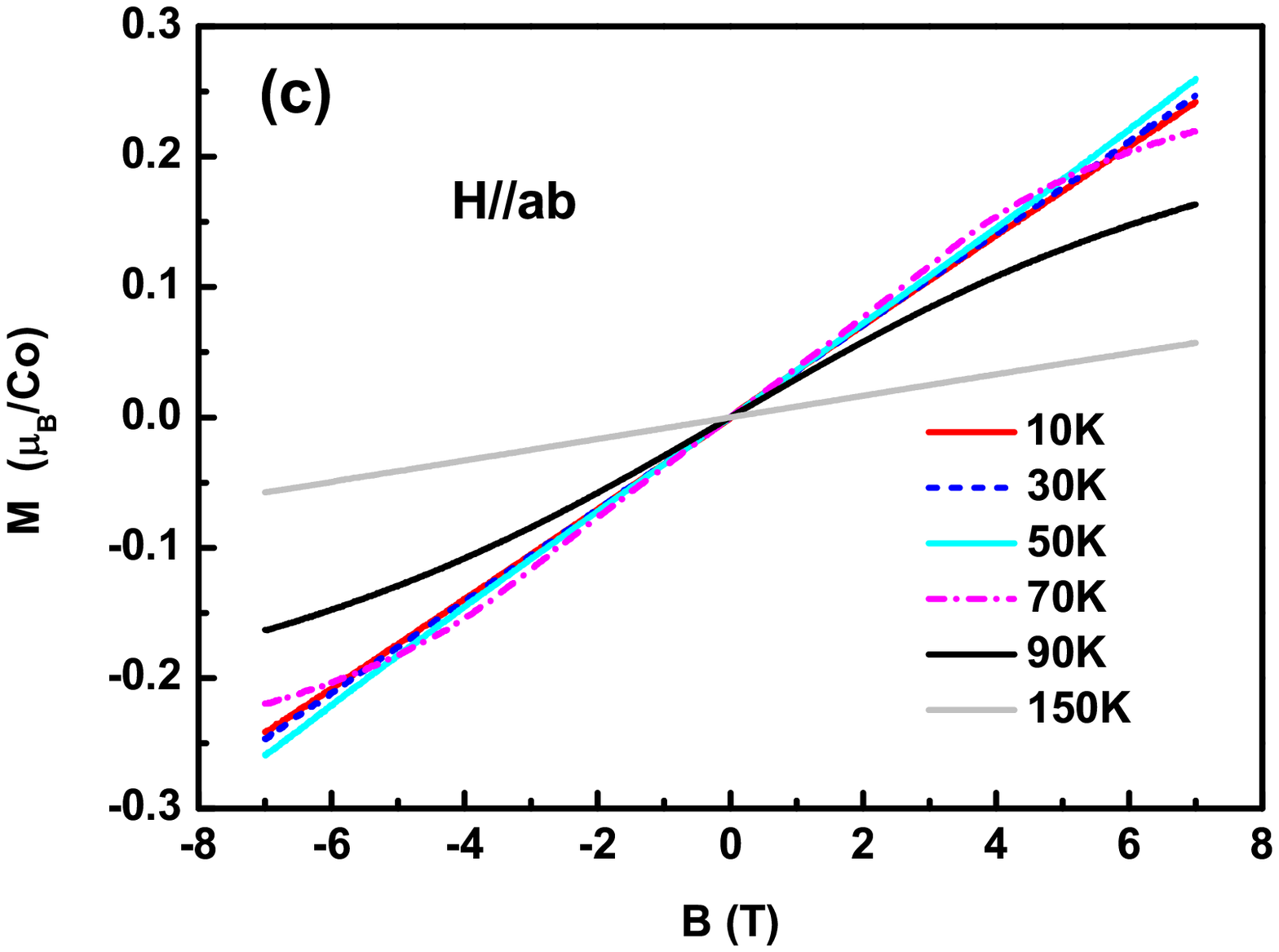}
\includegraphics[clip,width=2.3in]{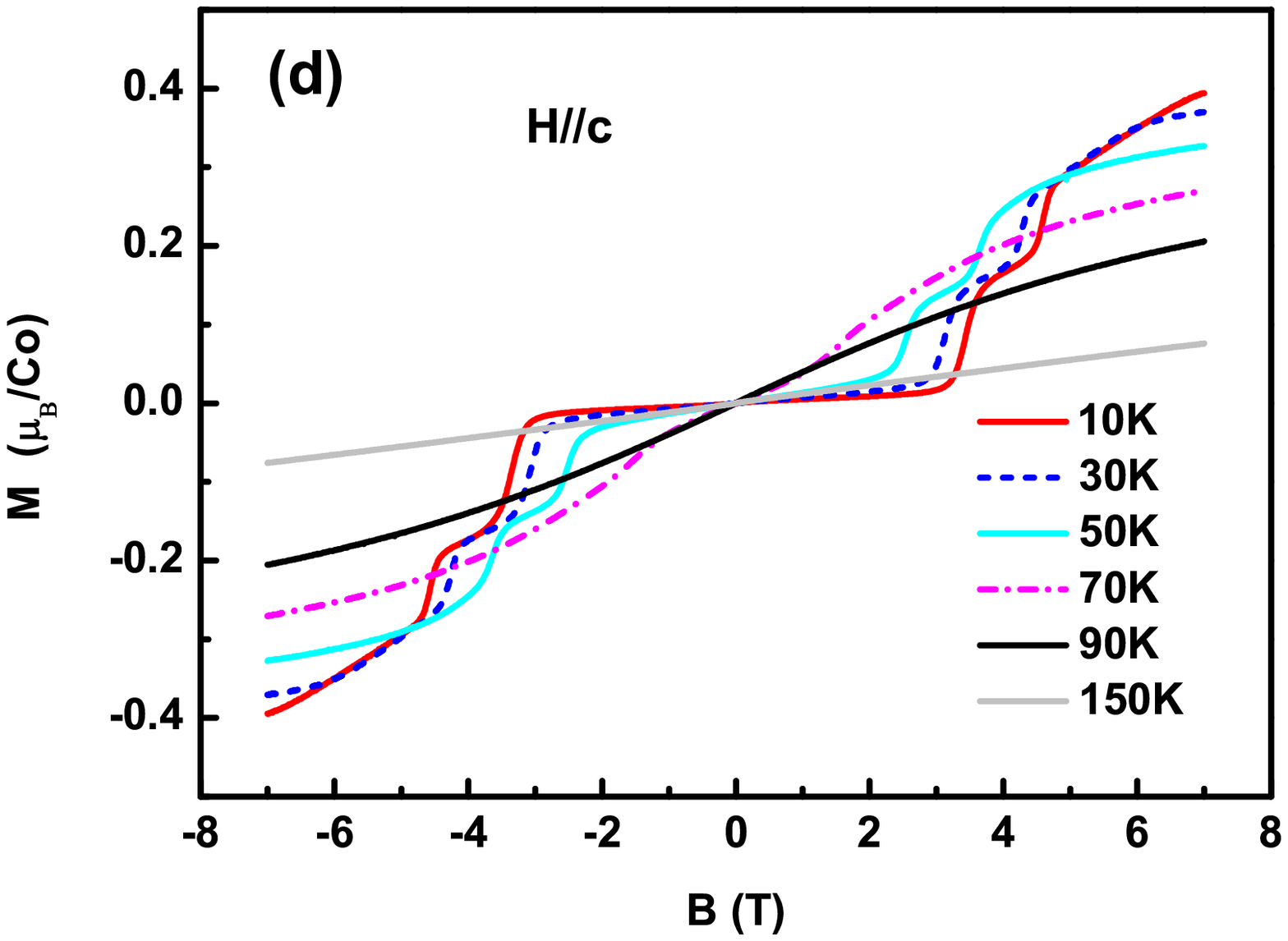}
\includegraphics[clip,width=2.3in]{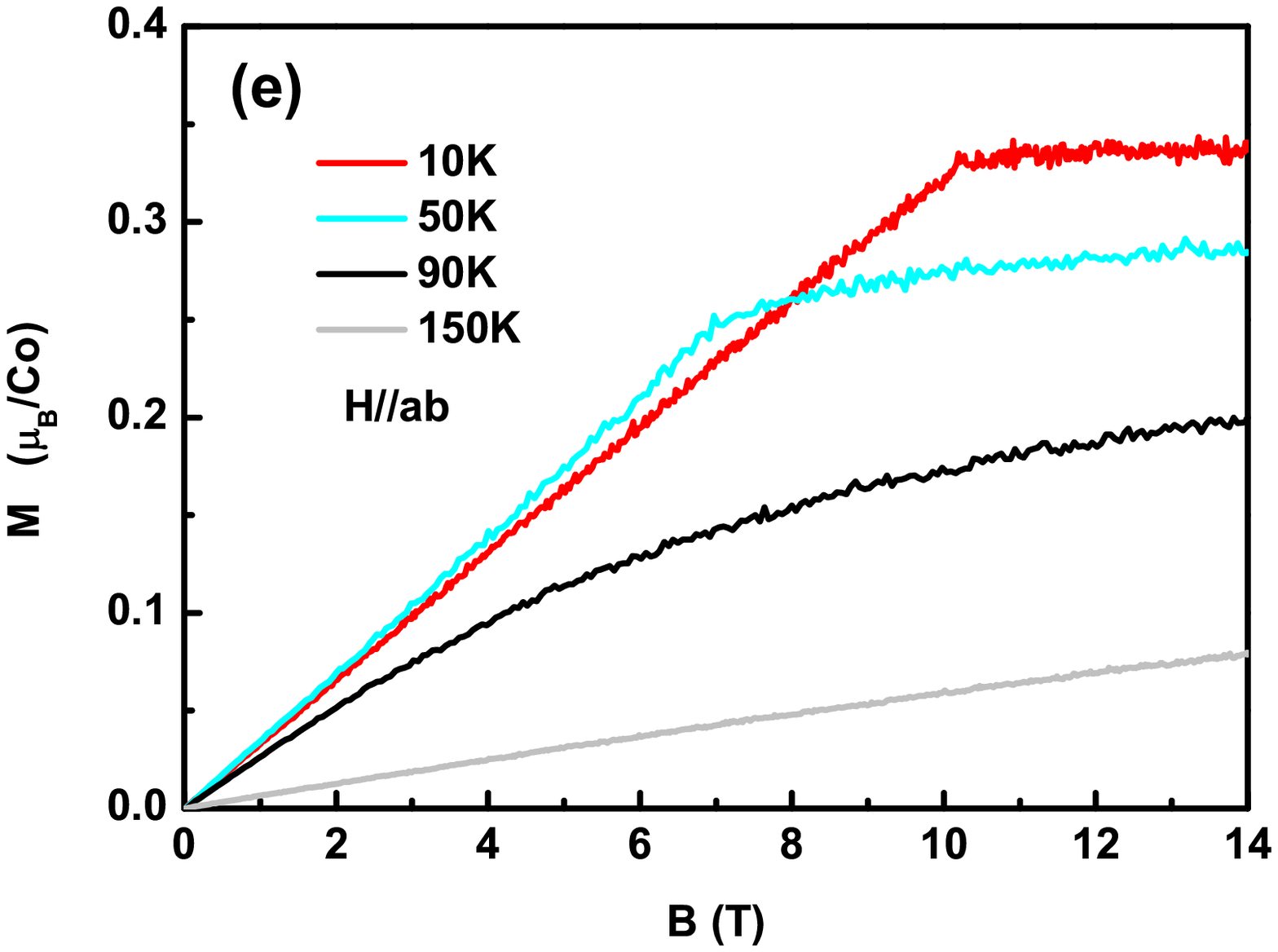}
\includegraphics[clip,width=2.3in]{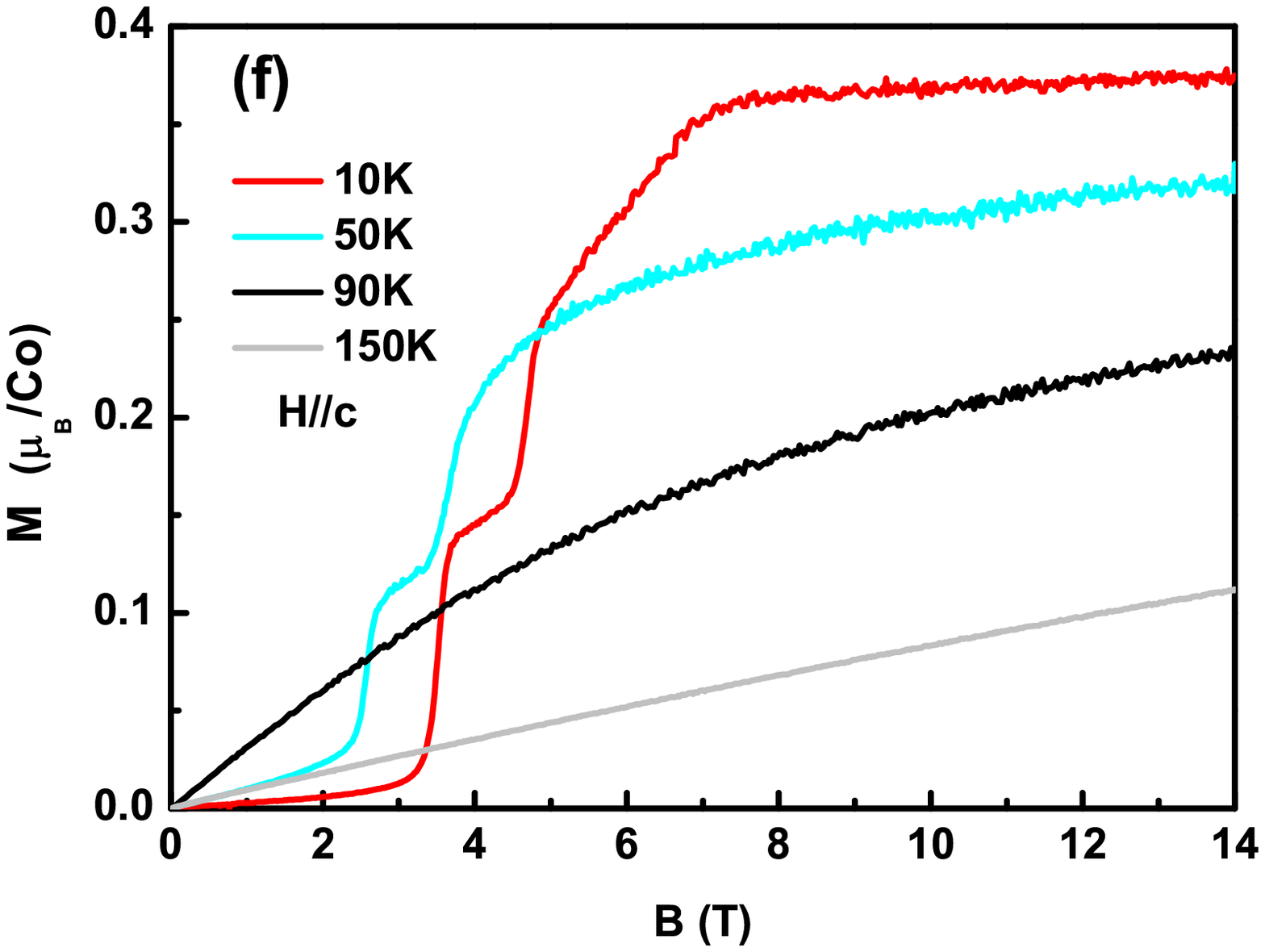}
\caption{(color online) (a) and (b) Susceptibilities with
\emph{\textbf{H} $\parallel$ ab} and \emph{ \textbf{H} $\parallel$
c} in different fields. (c) and (d) Magnetization as a function of
field from -7T to 7T below and above \emph{T$_{N}$} with
\emph{\textbf{H} $\parallel$ ab} and \emph{\textbf{H} $\parallel$
c}. (e) and (f) Magnetization as a function of field from 0 to 14T
below and above \emph{T$_{N}$} with \emph{\textbf{H} $\parallel$ ab}
and \emph{\textbf{H} $\parallel$ c}.}

\end{figure*}

Temperature dependent specific heat is shown in Fig. 3. Although the
contribution of phonon specific heat is dominant, we can observe a
weak peak locating around 76K clearly. It gives another evidence for
a bulk long-range antiferromagnetic transition. The fit to low
temperature specific heat data is shown in the inset of Fig. 3. A
good linear T$^{2}$-dependent behavior indicates that the low
temperature specific heat is mainly contributed by electrons and
phonons. The fit yields the electronic coefficient
\emph{$\gamma$}=30 mJ/K$^{2}$/mol CaCo$_{2}$As$_{2}$ or 15
mJ/K$^{2}$/mol Co atom. The value, which is much larger than that of
iron pnictide parents,\cite{bire,bire2} suggests a high density of
states (DOS) at the Fermi level. LDA calculation for
BaCo$_{2}$As$_{2}$ reveals that electronic DOS at the Fermi level is
already large enough to lead to a mean-field stoner instability
toward in-plane ferromagnetism.\cite{Co5} The large electronic
specific heat coefficient of CaCo$_{2}$As$_{2}$ may give us a clue
to understand the in-plane ferromagnetism in its antiferromagnetic
state.

Magnetic susceptibilities measured in different fields are shown in
Fig. 4(a) and 4(b). \emph{$\chi$$_{ab}$(T)} (\emph{$\chi$}(\emph{T})
with \emph{\textbf{H}}$\parallel$\emph{ab}) reveals a rather weak
field dependence up to 60 kOe. However, \emph{$\chi$$_{c}$(T)}
(\emph{$\chi$}(\emph{T}) with \emph{\textbf{H}}$\parallel$\emph{c})
(Fig. 4(b)) shows strong field dependent behaviors. The peak of
\emph{$\chi$$_{c}$(T)} shifts to low temperature with increasing
magnetic field. This is a characteristic feature of antiferromagntic
transition. Below 20 kOe, \emph{$\chi$$_{c}$(T)} reveals a
well-defined antiferromagnetic transition. However, at
\emph{\textbf{H}}=30 kOe, a shoulder begins to appear at 50K. With
increasing \emph{\textbf{H}}, the low temperature parts of
\emph{$\chi$$_{c}$(T)} are elevated and a plateau begins to form. At
50 kOe, \emph{$\chi$$_{c}$(T)} shows a large plateau below 40K,
which indicates that the antiferomagnetic ordering state is heavily
disturbed by the applied field. With further increasing magnetic
field, the low temperature plateau of \emph{$\chi$$_{c}$(T)} is
gradually suppressed . At \emph{\textbf{H}}=70 kOe, the plateau
could not be seen clearly.

To understand these peculiar behaviors of \emph{$\chi$$_{c}$(T)}, we
collected the magnetization data measured as a function of field
above and below \emph{T$_{N}$}. Figure. 4(c) shows
\emph{\textbf{M}(\textbf{H})} with \emph{\textbf{H}$\parallel$ab} in
the range of -7T to 7T. \emph{\textbf{M}$_{ab}$}(10K)
(\textbf{\emph{M}}(\emph{\textbf{H}},10K) with
\emph{\textbf{H}}$\parallel$\emph{ab}) and
\emph{\textbf{M}$_{ab}$}(150K) exhibit linear increase behavior as a
function of applied magnetic field.
\emph{\textbf{M}$_{ab}$(\textbf{H})} at 70K and 90K deviate from
linear behavior slightly. This anomaly originates from the strong
magnetic fluctuation around phase transition temperature. Figure.
4(d) gives the results with \emph{\textbf{H}$\parallel$c}. Different
from \emph{\textbf{H}$\parallel$ab}, \emph{\textbf{M}$_{c}$}(10K)
(\textbf{\emph{M}}(\emph{\textbf{H}},10K) with
\emph{\textbf{H}}$\parallel$\emph{c}) undergoes two steep
magnetization jumps at $\mu$$_{0}$\emph{\textbf{H}$_{c1}$}=3.5T and
$\mu$$_{0}$\emph{\textbf{H}$_{c2}$}=4.7T. The first jump exhibits a
notable hysteresis in the \emph{\textbf{M}}-\emph{\textbf{H}} curve
as shown in Fig. 5. However, the second one at 4.7T hardly shows a
hysteresis. From 5T to 7T, no hysteresis can be observed in
\emph{\textbf{M}$_{c}$}(10K). With increasing temperature, the two
magnetization jumps become less pronounced and disappear above 70K.
\emph{\textbf{M}$_{c}$(\textbf{H})} at 150K shows a good linear
behavior as a function of magnetic field, indicating a typical
paramagnetic response in paramagnetic states. We noticed that
similar steep jump behaviors have been observed in many
antiferromagnets, such as CuCl$_{2}$$\cdot$2H$_{2}$O,\cite{3D1}
Cu$_{2}$MnSnS$_{4}$,\cite{3D3} BaCu$_{2}$Si$_{2}$O$_{7}$,\cite{1D1}
Na$_{0.85}$CoO$_{2}$\cite{NCO} and
${\beta}$-Cu$_{2}$V$_{2}$O$_{7}$.\cite{1D2} A natural explanation to
the steep magnetization jump behaviors in antiferromagnet is a
spin-flop transition. To yield more information on the jumps, we
performed magnetization measurements up to 14T.

 Figure. 4(e) and 4(f) show \emph{\textbf{M}$_{ab}$(\textbf{H})} and
\emph{\textbf{M}$_{c}$(\textbf{H})} up to 14T at different
temperatures. At 10K, \emph{\textbf{M}$_{ab}$(\textbf{H})} and
\emph{\textbf{M}$_{c}$(\textbf{H})} display moment saturation
behaviors at 10.2T and 7.6T respectively. The saturation moments are
0.33\emph{$\mu$$_{B}$} per Co for \emph{\textbf{H}$\parallel$ab} and
0.37\emph{$\mu$$_{B}$} per Co for \emph{\textbf{H}$\parallel$c}.
These values are much smaller than the effective moment per Co atom
obtained from Curie-Weiss fits to the susceptibilities, indicative
of an itinerant magnetism in CaCo$_{2}$As$_{2}$. With increasing
temperature, the saturation behaviors are weakened and finally
disappear above \emph{T$_{N}$}.

\begin{figure}[b]
\scalebox{0.45} {\includegraphics [bb=300 30 10cm 12cm]{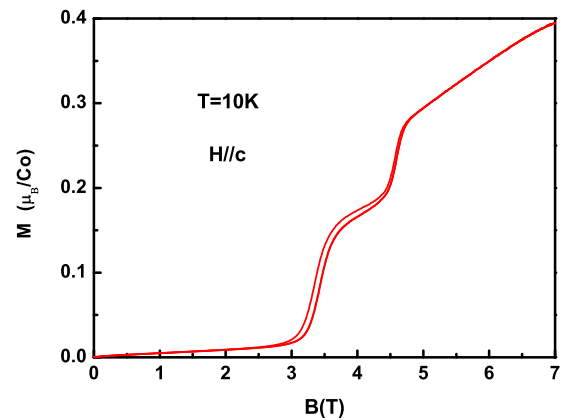}}
\caption{(color online) \emph{\textbf{M}$_{c}$} versus
\emph{\textbf{H}} measured up to 7T at 10K with increasing and then
decreasing \emph{\textbf{H}} }
\end{figure}

It is well known that spin-flop transition can be induced by a
moderate magnetic field in an uniaxial antiferromagnet with low
anisotropy. The first jump displays a notable hysteresis with
increasing and then decreasing \emph{\textbf{H}}, indicating a
first-order phase transition. We infer that this jump is ascribed to
the traditional spin-flop transition which has been observed in many
uniaxial antiferromagnets. The possible magnetic structures of
CaCo$_{2}$As$_{2}$ in this spin-flop phase are shown in Fig. 6(c)
and 6(d). At 4.7T, another sudden jump occurs in
\emph{\textbf{M}$_{c}$(\textbf{H})} at 10K. This jump is unexpected
from a simple uniaxial antiferromagnet. The steep increase of
magnetization means that the magnetic structure of
CaCo$_{2}$As$_{2}$ undergoes a sudden change. Different from the
first jump, the second one exhibits much weaker hysteresis behavior.
We can not give a detailed description about this spin-flop
transition because exact magnetic structure and magnetic interaction
in CaCo$_{2}$As$_{2}$ can not be obtained from static susceptibility
and magnetization data. We think further neutron experiments are
needed to settle this issue. For \emph{\textbf{H}$\parallel$ab},
behaviors of moments responding to the external field are much
simpler than that of \emph{\textbf{H}$\parallel$c}. The balance
between Zeeman energy, antiferromagnetic coupling energy and
magnetocrystalline anisotropic energy lead that the magnetic moments
are gradually rotated to \emph{ab} plane. Above 10.2T, all moments
lie in \emph{ab} plane and are ordered ferromagnetically as shown in
Fig. 6(f).

\begin{figure}[t]
\scalebox{0.35} {\includegraphics [bb=300 25 8cm 18cm]{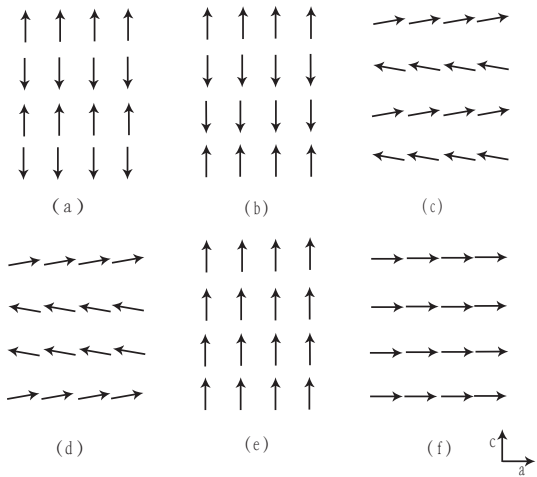}}
\caption{(a) and (b) Two possible magnetic structures in
CaCo$_{2}$As$_{2}$. (c) and (d) Two possible spin-flop phases in
CaCo$_{2}$As$_{2}$. (e) Arrangement of Co moments in large field
with  \emph{\textbf{H}$\parallel$c}. (f) Arrangement of Co moments
in large field with \emph{\textbf{H}$\parallel$ab}.}
\end{figure}

These interesting magnetic phenomena give us a chance to estimate
the antiferromagnetic exchange coupling energy along the \emph{c}
axis and the magnetocrystalline anisotropic energy in
CaCo$_{2}$As$_{2}$. The behavior of
\emph{\textbf{M}$_{c}$(\textbf{H})} in magnetic states is mainly
determined by interlayer antiferromagnetic exchange coupling energy
\emph{E$_{c}$=$\sum$$_{ij}$J$_{c}$\textbf{S}$_{i}$$\cdot$\textbf{S}$_{j}$}
and Zeeman energy
\emph{E$_{z}$=-$\sum$$_{i}$\textbf{m}$_{i}$$\cdot$\textbf{B}} and
magnetocrystalline anisotropic energy. \emph{J$_{c}$} is coupling
const. \emph{\textbf{S}} and \emph{\textbf{m}} is the spin and
moment of Co ions. The saturation behavior above 7.6T in
\emph{\textbf{M}$_{c}$}(10K) manifests that the moments of Co atoms
are all aligned ferromagnetically along the \emph{c} axis as shown
in Fig. 6(e). At 7.6T, we assume that the energy gain of Zeeman
interaction can just overcome the energy cost induced by the
antiferromagnetic exchange coupling when the moments are flipped. In
this method, we find a simply approximate relation :
\emph{E$_{z}$=-2E$_{c}$} at $\mu$$_{0}$\emph{\textbf{H}}=7.6T for
\emph{\textbf{H}$\parallel$c}. Taking the estimated value of
\emph{m} $\sim$ 0.4\emph{$\mu$$_{B}$} from the magnetization
saturation and assuming \emph{g}=2, we estimate \emph{J$_{c}$}
$\sim$ 2 meV. Neutron scattering data reveal that \emph{J$_{c}$} in
122 parent compounds of iron-based superconductors varies from 1 meV
to 10 meV.\cite{neutron} Our estimation about \emph{J$_{c}$} in
CaCo$_{2}$As$_{2}$ exhibits the same order of magnitude with 122
parent compounds.

\emph{J$_{c}$} in  CaCo$_{2}$As$_{2}$ is much lower than that of
some other antiferromagnets, such as Na$_{0.85}$CoO$_{2}$, whose
\emph{J$_{c}$} determined by neutron diffraction is 12.2
meV.\cite{zhongzi} In Na$_{0.85}$CoO$_{2}$, a spin-flop transition
was observed, which is very similar to the first spin-flop
transition in CaCo$_{2}$As$_{2}$.\cite{NCO} But up to 14T, the
saturated phenomenon of \emph{\textbf{M}$_{c}$}(5K) in
Na$_{0.85}$CoO$_{2}$ had not been observed. This means that
\emph{J$_{c}$} in Na$_{0.85}$CoO$_{2}$ is too high for 14T to induce
the similar magnetization saturation which occurs in
CaCo$_{2}$As$_{2}$.

The magnetization saturation behaviors for
\emph{\textbf{H}$\parallel$ab} and \emph{\textbf{H}$\parallel$c}
provide some information on the magnetocrystalline anisotropy. That
the moments of Co atoms are aligned along the \emph{c} axis at zero
field indicates that it will cost more energy when the moments lie
in \emph{ab} plane. To achieve the magnetic state as shown in Fig.
6(e), Zeeman energy must overcome the energy cost caused by magnetic
exchange interaction when the moments are flipped. However, to
achieve the magnetic state as shown in Fig. 6(f) with
\emph{\textbf{H}$\parallel$ab}, Zeeman energy must overcome extra
energy cost induced by magnetocrystalline anisotropic energy. This
magnetocrystalline anisotropic energy can be estimated through the
difference between the saturation fields of
\emph{\textbf{H}$\parallel$ab} and \emph{\textbf{H}$\parallel$c}. In
this method, the magnetocrystalline anisotropic energy is estimated
to be about 3.76 $\times$ 10$^{5}$ erg/g.

\section{\label{sec:level2}SUMMARY}

In summary, we have investigated transport and magnetic properties
of single-crystalline CaCo$_{2}$As$_{2}$ by means of resistivity,
heat capacity, magnetic susceptibility and magnetization
measurements. Our results reveal that CaCo$_{2}$As$_{2}$ undergoes
an antiferromagnetical transition at \emph{T$_{N}$}=76K. The
estimated value of ordered moment on Co atom is about
0.4\emph{$\mu$$_{B}$}. Two successive spin-flop transitions have
been observed at $\mu$$_{0}$\emph{\textbf{H}$_{c1}$}=3.5T and
$\mu$$_{0}$\emph{\textbf{H}$_{c2}$}=4.7T in
\emph{\textbf{M}$_{c}$}(10K). Our analyses indicate that
antiferromagnetic coupling between \emph{ab} plane is rather weak.
The interlayer antiferromagntic coupling constant and the
magnetocrystalline anisotropic energy are estimated to be about 2
meV and 3.76 $\times$ 10$^{5}$ erg/g respectively.

\begin{center}
\small{\textbf{ACKNOWLEDGMENTS}}
\end{center}
This work was supported by the National Science Foundation of China
(10834013, 11074291) and the 973 project of the Ministry of Science
and Technology of China (2011CB921701)


\begin{references}
\bibitem{RMP1} C. J. Gorter,  Rev. Mod. Phys  \textbf{22}, 277 (1953).
\bibitem{LiLun1} A. N  Bodganov, A. V. Zhuravlev, U. K. R\"{o}{\ss}ler,  Phys. Rev. B \textbf{75}, 094425 (2007).
\bibitem{3D1} C. J. Gorter,  Rev. Mod. Phys  \textbf{25}, 332 (1953).
\bibitem{3D2-LSMO} U. Welp, A. Berger, D. J. Miller, V. K. Vlasko-Vlasov, K. E. Gray, and J. F, Mitchell, Phys. Rev. Lett \textbf{83}, 4180 (1999).
\bibitem{3D3} T. Fries, Y. Shapira, Fernando Palacio, M. Carmen Mor\'{o}n, Garry J. Mclntyre, R. Kershaw, A.Wold, and E. J Mcniff. Jr, Phys. Rev. B \textbf{56}, 5424  (1997).
\bibitem{3D4} Y. Shapira, and S. Foner, Phys. Rev \textbf{170}, 503 (1968)
\bibitem{1D1} I. Tsukada, J. Takeya, T. Masuda, and K. Uchinokura, Phys. Rev. Lett. \textbf{87}, 127203 (2001).
\bibitem{1D2} Zhangzhe He, and Yutaka Ueda,  Phys. Rev. B \textbf{77}, 052402 (2008).
\bibitem{1D3} D. A. Zocco, J. J. Hamlin, T. A. Sayles, M. B. Maple, J. H. Chu and I. R. Fisher, Phys. Rev. B \textbf{79}, 134428 (2009).
\bibitem{2D1} R. W. Wang, D. L. Mills, Eric E. Fullerton, J. E. Mattson, and S. D. Bader, Phys. Rev. Lett \textbf{72}, 920, (1994)
\bibitem{Fe1} Y. Kamihara, T. Watanabe, M. Hirano, and H. Hosono, J. Am. Chem. Soc \textbf{130}, 3296, (2008)
\bibitem{Fe2} G. F. Chen, Z. Li, D. Wu, G. Li, W. Z. Hu, J. Dong, P. Zheng, J. L. Luo, and N. L. Wang, Phys. Rev. Lett \textbf{100}, 247002, (2008)
\bibitem{Fe3} Marianne Rotter, Marcus Tegel, and Dirk Johrendt, Phys. Rev. B \textbf{78}, 020530\textbf{R}, (2008)
\bibitem{Fe4} Z. A. Ren, W. Lu, J. Yang, W. Yi, X. L. Shen, Z. C. Li, G. C. Che, X. L. Dong, L. L. Sun, F. Zhou, and Z. X. Zhao, Chin. Phys. Lett \textbf{25}, 2215, (2010)
\bibitem{Co1} Hiroto Ohta, and Kazuyoshi Yoshimura, Phys. Rev. B \textbf{80}, 184409, (2009)
\bibitem{Co2} V. P. S. Awana, I. Nowik, Anand. Pal, K. Yamaura, E. Takayama-Muromachi, and I. Felner,  Phys. Rev. B \textbf{81}, 212501, (2010)
\bibitem{Co3} Michael A. McGuire, Delphine J. Gout, V. Ovidiu Garlea, Athena S. Sefat, Brian C. Sales, and David Mandrus, Phys. Rev. B \textbf{81}, 104405, (2009)
\bibitem{Co4} Andrea Marcinkova, David A. M. Grist, Irene Margiolaki, Thomas C. Hansen, Serena Margadonna, and Jan-Willem G. Bos, Phys. Rev. B \textbf{81}, 064511, (2010)
\bibitem{Co5} A. S. Sefat, D. J. Singh, R. Jin, M. A. McGuire, B. C. Sales, and D. Mandrus, Phys. Rev. B \textbf{79}, 024512, (2009)
\bibitem{flux} X. F. Wang, T. Wu, G. Wu, H. Chen, Y. L. Xie, J. J. Ying, Y. J. Yan, R. H. Liu, and X. H. Chen,  Phys. Rev. Lett \textbf{102}, 117005, (2009)
\bibitem{P1}  M. Reehuis, W. Jeitschko, G. Kotzyba, B. Zimmer, X. Hu, J. Alloys. comp \textbf{266}, (1998), 54
\bibitem{P2}  M. Reehuis, P. J. Brown, W. Jeitschko, M. H. M\"{o}ller, and T. Vomhof, J. Phys. Chem. Solids \textbf{54}, 469, (1993)
\bibitem{bire} G. F. Chen, Z. Li, J. Dong, W. Z. Hu, X. D. Zhang, X. H. Song, P. Zheng, N. L. Wang, and J. L. Luo, Phys. Rev. B \textbf{78}, 224512, (2008)
\bibitem{bire2} G. F. Chen, W. Z. Hu, J. L. Luo, and N. L. Wang, Phys, Rev. Lett \textbf{102}, 227004, (2009)
\bibitem{neutron} D. C. Johnston, Advances in Physics  \textbf{59},
803, (2010)

\bibitem{NCO} J. L. Luo, N. L. Wang, G. T. Liu, D. Wu, X. N. Jing, F. Hu, and T. Xiang, Phys. Rev. Lett \textbf{93}, 187203, (2004)
\bibitem{zhongzi} L. M. Helme, A. T. Boothroyd, R. Coldea, D. Prabhakaran, A. Stunault, G. J. Mcintyre, and N.Kernavanois, Phys. Rev. B \textbf{73}, 054405, (2006)









\end{references}
\end{document}